\documentclass[journal]{IEEEtran}
\usepackage{graphicx}% http://ctan.org/pkg/graphicx
\usepackage[utf8]{inputenc}
\usepackage{textcomp}
\usepackage{gensymb}
\usepackage{multirow}
 \usepackage{svg}
 \usepackage{float}
\usepackage{amsmath,amssymb}
\usepackage{tikz}
\usetikzlibrary{matrix}
\usepackage{verbatim}
\usepackage{cleveref}
\usepackage{tabularx}
\usepackage{algorithm,algpseudocode}

\usetikzlibrary{positioning,arrows,arrows.meta}
\usepackage{subcaption}
\usepackage[skip=2pt,font=footnotesize]{caption}

\ifCLASSINFOpdf
\else
\fi

\begin{document}
\title{Deep Learning Applied to Beamforming \\ in Synthetic Aperture Ultrasound}

% author names and IEEE memberships
\author{Nissim Peretz,
        and~Arie~Feuer,~\IEEEmembership{Life~Fellow,~IEEE}% <-this % stops a space
\thanks{N. Peretz is with the Department of Electrical Engineering, Technion-Israel Institute of Technology, Haifa 3200003, Israel (e-mail: nissimp@campus.technion.ac.il).}% <-this % stops a space
\thanks{A. Feuer is with the Department of Electrical Engineering, Technion-Israel Institute of Technology, Haifa 3200003, Israel (e-mail: feuer@ee.technion.ac.il).}% <-this % stops a space
}

% The paper headers
%\markboth{Journal of \LaTeX\ Class Files,~Vol.~14, No.~8, August~2015}{Shell \MakeLowercase{\textit{et al.}}: Bare Demo of IEEEtran.cls for IEEE Journals}

% If you want to put a publisher's ID mark on the page you can do it like
% this:
%\IEEEpubid{0000--0000/00\$00.00~\copyright~2015 IEEE}
% Remember, if you use this you must call \IEEEpubidadjcol in the second column for its text to clear the IEEEpubid mark.

% make the title area
\maketitle

\begin{abstract}
Deep learning methods can be found in many medical imaging applications. Recently, those methods were applied directly to the RF ultrasound multi-channel data to enhance the quality of the reconstructed images. In this paper, we apply a deep neural network to medical ultrasound imaging in the beamforming stage. Specifically, we train the network using simulated multi-channel data from two arrays with different sizes, using a variety of direction of arrival (DOA) angles, and test its generalization performance on real cardiac data. We demonstrate that our method can be used to improve image quality over standard methods, both in terms of resolution and contrast. Alternatively, it can be used to reduce the number of elements in the array, while maintaining the image quality. The utility of our method is demonstrated on both simulated and real data.
\end{abstract}


% For peer review papers, you can put extra information on the cover
% page as needed:
% \ifCLASSOPTIONpeerreview
% \begin{center} \bfseries EDICS Category: 3-BBND \end{center}
% \fi
%
% For peerreview papers, this IEEEtran command inserts a page break and
% creates the second title. It will be ignored for other modes.
\IEEEpeerreviewmaketitle

\section{Introduction}

\IEEEPARstart{B}{eamforming} is a crucial task in a variety of fields
including, radar, sonar, acoustics and wireless communication \cite%
{van1988beamforming}. While there are several different approaches in the
literature, the Delay-And-Sum (DAS) beamforming is the most common method in
medical ultrasound imaging. An image is formed by transmitting a narrow beam
in several scanning angles, and dynamically delaying and summing the
received signals from all channels. The images produced by DAS are typically
degraded by sidelobes artifacts, resulting from the processing applied on
the data collected. Sidelobes are considered as interferences that mask the
desired signal, reducing the contrast-to-noise ratio (CNR). The sidelobe
level of the DAS beamformer can be controlled using \textit{aperture
apodization} (amplitude weighting of the elements across the aperture),
resulting in increased contrast at the expense of resolution. Thus, the
weights used in the beamforming process are designed to trade off sidelobes
reduction for lateral resolution.

The conventional ultrasound scanning method transmits pulses in one
direction at a time. The signals received by the transducer, are used to
reconstruct a part of an image corresponding to the scan line. This set of
lines is then interpolated onto a Cartesian grid, and a single frame can be
created. The process is repeated sequentially to create the next frames.
Consequently, the acquisition time is limited by the speed of sound. Another
limiting factor in conventional ultrasound imaging is the single transmit
focus, where the transmit beam is only focused at one depth. This can be
solved by image compounding \cite{cheng2006extended} using several beams
with different transmit focus, resulting in a very large depth of focus, but
the frame rate is then correspondingly decreased. Therefore, there is much
interest in searching alternatives to the conventional methods, where the
single transmit focusing and frame rate constraints can be alleviated. An
alternative is to use \textit{Synthetic Aperture} imaging techniques.

In the first technique, known as \textit{Synthetic Aperture} (SA) ultrasound 
\cite{jensen2006synthetic}, only a single array element transmits and
receives each time. All array elements transmit/receive sequentially one at
a time and the received echoes are sampled and stored. SA increases the
frame rate, but the main limitation of the technique is the low SNR and poor
contrast resolution (see \cite{jensen2006synthetic}).

In the second technique, known as \textit{Synthetic Transmit Aperture} (STA)
ultrasound \cite{chiao1997sparse}, a single element transmits and all
elements receive the reflected signals simultaneously. The SNR is increased compared to
SA, since the image reconstruction uses a larger amount of data collected by
all elements. This multistatic approach produces the image in a tomographic
manner in which the numerous transmit/receive angles improve the resolution
and contrast of the reconstructed image. The main limitation of the STA
technique is the very large amount of data that needs to be stored in memory.

Norton \cite{norton1980reconstruction} has modeled the SA method, under some simplifying
assumptions, the process of generating the data from an imaged object as a
forward mapping. Hence, generating the (unknown) image from the accumulated
data is viewed as an inverse mapping. He proceeded to analytically generate
a closed form expression for this inverse mapping.

When the data is fed into the inverse system, it is convolved with the
appropriate impulse response to produce the imaged object. DAS beamforming
is a linear approximate solution to the inverse system, which is not
accurate in the presence of multiple reflections and multiple varying
acoustic properties.

In recent years, the field of Deep Learning has gained a lot of momentum in
solving such inverse problems in imaging \cite{lucas2018using}. Inspired by
the feasibility and success of this field, we propose in this work to
replace the traditional DAS beamforming process with an alternative using
DNN as part of the scanned image forming.

\subsection{Related Work}

A variety of ultrasound reconstruction approaches utilizing deep learning
have been proposed in the literature.

Luchies and Byram \cite{luchies2018deep} investigated the use of multiple
networks for reducing off-axis single point scattering in ultrasound images,
processing the data in the frequency domain using the short-time Fourier
transform. For each frequency bin, the array samples are processed by a
different network. The data was transformed back into the time domain using
the inverse Fourier-transform and summed across the array to create a signal
that corresponds to a single depth of channel data. In \cite%
{luchies2019training}, the authors aim to extend their previous work, by training
multiple networks using multiple point target reflections instead of single
point target reflection.

Gasse et al. \cite{gasse2017high} proposed to exploit neural networks to
reduce the number of emitted transmissions in coherent plane-wave
compounding. A convolutional neural network (CNN) was trained to reconstruct
high-quality images, based on a small number of transmissions. The authors
in \cite{nair2018fully} posed the beamforming process in ultrasound as a
segmentation problem and applied a CNN to segment cyst phantoms.

In \cite{senouf2018high}, \cite{vedula2018high}, the authors suggested
exploiting CNNs to reconstruct high-quality images acquired through
high-frame rate ultrasound techniques, such as multi-line acquisition (MLA)
and multi-line transmission (MLT). They propose to train a network that
takes MLA/MLT channel data as an input, and reconstructs images such that it
mimics the operation of a single-line acquisition (SLA). In \cite%
{vedula2018learning}, they proposed to learn the parameters of the forward
model simultaneously with the image reconstruction process, that is, to
jointly learn the end-to-end transmit and receive beamformers.

In \cite{simson2018deep}, the authors proposed a method to reconstruct
high-quality ultrasound images in real-time on sub-sampled scan lines data.
The network learns a mapping from a sub-sampled collection of scan lines to
high-quality images, which were generated offline with a minimum variance
beamforming technique.

Luijten et al. \cite{luijten2019deep} proposed to approximate the
time-consuming Eigen Based Minimum Variance beamformer using a CNN, which
enable a real-time implementation of adaptive beamforming in plane-wave
imaging. In \cite{perdios2018deep}, the authors proposed to use a CNN to
learn a non-linear mapping from the low-quality images, reconstructed from a
single plane-wave, to the high-quality images, reconstructed from multiple
plane-waves.

Reviewing all the mentioned works, it appears that none of the results
relates to the aperture size and number sensors in the probe which is one of
the main attributes of our work. Furthermore, \ the Synthetic Aperture
imaging techniques are not addressed. Concluding this review, it appears
that the potential benefits of applying deep learning in ultrasound medical
imaging are yet to be exhausted.

\subsection{Contributions}
The image reconstruction method we present here, DNN Beamforming (DNNB),
combines two independent novel ideas. One, that a DNN can be trained to use
scanning data generated by a given array size to emulate scanning data from
a larger array, and two, that a DNN can be trained to suppress sidelobe
artifacts when applied with DAS. While initially we have trained two DNNs to
achieve the above, by using synthetic data for the network training, we managed
to combine the two into a single network. The experiments we present here are
all using a single DNN trained for both tasks.

We applied our method, DNNB, on three major scanning techniques: \textit{%
Synthetic Aperture} (SA), \textit{Synthetic Transmit Aperture} (STA) and
conventional \textit{Phased Array} (PA), using a properly trained neural network
as part of the beamforming process in each. We demonstrate that our method
clearly outperforms the DAS approach for all three techniques in terms of
the resulting image quality. The resulting images are almost free of DAS
induced sidelobe artifacts and have significantly improved lateral
resolution compared to DAS images using the \emph{same aperture}.

We visualize three potential scenarios for applying our method. One,
improving the reconstructed image quality of a given scanning array size. Two,
using a smaller scanner, which is frequently desirable, while maintaining an
acceptable reconstructed image quality, and three, when data size is an
issue with a given size scanner, using full array size in transmission and
downsampling the received data - using only the data from a subset or
receiving elements.

\begin{figure}[t!]
\centering
\begin{tikzpicture}
        \draw[thick,->] (-2,0) -- (5.5,0) node[anchor=north west] {x};
        \draw[thick,->] (0,0) -- (0,4) node[anchor=south east] {z};
        \draw[thick,-] (4,0) arc (0:180:2.5cm);
        \draw[thick,->] (1.5,0) -- (2.95,2) node[anchor=south ] {};
        \draw[thick,-] (1.95,1) -- (1.95,1) node[anchor=south ] {$r$};
        \draw[thick,-] (1.5,0) -- (1.5,0) node[anchor=north ] {$x_0$};
        \draw[thick,-] (3.95,3) -- (3.95,3) node[anchor=south ] {$f(x,z)$};
        \draw[thick,-] (-0.95,-0.75) -- (-0.95,-0.75) node[anchor=south ] {Boundary};
        \draw[thick,-] (-0.95,3) -- (-0.95,3) node[anchor=south ] {Medium};
        \draw[thick,-] (4.5,1) -- (4.5,1) node[anchor=south ] {$C\left(x_{0}, r\right)$};
    \end{tikzpicture}
\caption{SA technique: circular wave front transmitted from $x_0$, and
received at $x_0$.}
\label{fig:fxz}
\end{figure}
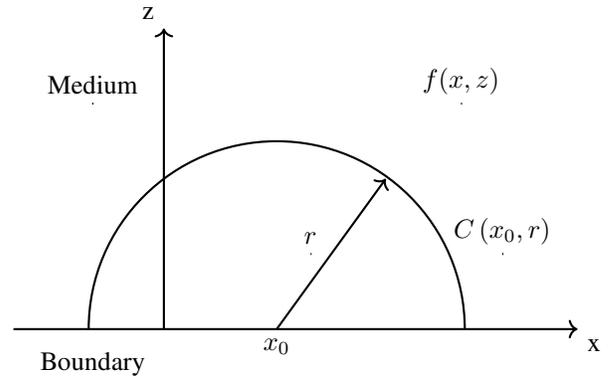

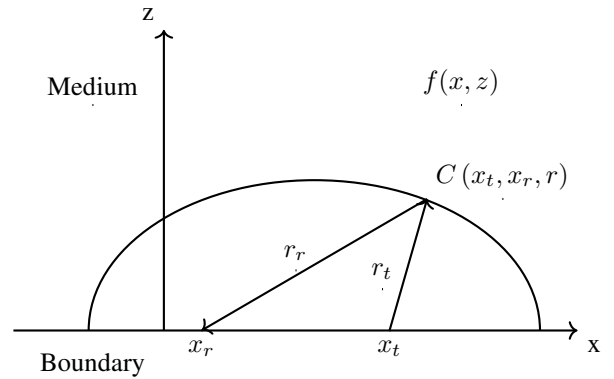
\begin{figure}[t!]
\centering
\begin{tikzpicture}
        \draw[thick,->] (-2,0) -- (5.5,0) node[anchor=north west] {x};
        \draw[thick,->] (0,0) -- (0,4) node[anchor=south east] {z};
        \draw[thick,-]  (-1,0) arc (-180:-360:3 and 2);
        \draw[thick,->] (3,0) -- (3.5,1.75) node[anchor=south ] {};
        \draw[thick,-] (2.9,0.55) -- (2.9,0.55) node[anchor=south ] {$r_t$};
        \draw[thick,-] (3,0) -- (3,0) node[anchor=north ] {$x_t$};
        \draw[thick,->] (3.5,1.75) -- (0.5,0) node[anchor=south ] {};
        \draw[thick,-] (0.5,0) -- (0.5,0) node[anchor=north ] {$x_r$};
        \draw[thick,-] (1.75,0.8) -- (1.75,0.8) node[anchor=south ] {$r_r$};
        \draw[thick,-] (3.95,3) -- (3.95,3) node[anchor=south ] {$f(x,z)$};
        \draw[thick,-] (-0.95,-0.75) -- (-0.95,-0.75) node[anchor=south ] {Boundary};
        \draw[thick,-] (-0.95,3) -- (-0.95,3) node[anchor=south ] {Medium};
        \draw[thick,-] (4.5,1.75) -- (4.5,1.75) node[anchor=south ] {$C\left(x_{t}, x_{r},r\right)$};
    \end{tikzpicture}
\caption{STA technique: circular wave front transmitted from $x_t$, and
received at $x_r$.}
\label{fig:fxz_STA}
\end{figure}

\begin{figure*}[t!]
\centering
\begin{tikzpicture}
  % CNN in
  \node[inner sep=0pt] (DNN_node0) at (-9.5,1) {(a)};
  \node[draw, text width=2cm,align=center] (DNN_node1) at (-7,1) {Simulated data};
  \node[draw, text width=2cm,align=center] (DNN_node2)
  at (-4,2) {Large Aperture};
  \node[draw, text width=2cm,align=center] (DNN_node3) at (-1,2) {Large Data};

  \node[draw, text width=2cm,align=center] (DNN_node22) at (-4,0) {Small Aperture};
  \node[draw, text width=2cm,align=center] (DNN_node23) at (-1,0) {Small Data};
  \node[draw, text width=2cm,align=center] (DNN_node25) at (2,0) {DNN};
    \node[draw, text width=2cm,align=center, dashed] (DNN_node26) at (5,0) {MSE Loss};
    
  % CNN out
  \draw[-{Triangle[scale=1.5]},draw=blue] (DNN_node1) to[in=180,out=0] (DNN_node2);
  \draw[-{Triangle[scale=1.5]},draw=blue] (DNN_node1) to[in=180,out=0] (DNN_node22);
  \draw[-{Triangle[scale=1.5]},draw=blue] (DNN_node2) to (DNN_node3);
  \draw[-{Triangle[scale=1.5]},draw=blue] (DNN_node22) to (DNN_node23);
  \draw[-{Triangle[scale=1.5]},draw=blue] (DNN_node23) to[in=180,out=0] (DNN_node25);
    \draw[-{Triangle[scale=1.5]},draw=blue] (DNN_node3) to[in=90,out=0] (DNN_node26);
    \draw[-{Triangle[scale=1.5]},draw=blue] (DNN_node25) to[in=180,out=0] (DNN_node26);
    \draw[-{Triangle[scale=1.5]},draw=blue, dashed] (DNN_node26) to[in=270,out=0] (DNN_node25);
    
    \end{tikzpicture}

\begin{tikzpicture}
  % CNN in
  \node[inner sep=0pt] (DNN_node0) at (-8,0) {(b)};
  \node[inner sep=0pt] (DNN_node1) at (-6.5,0) {\includegraphics[width=0.1\textwidth]{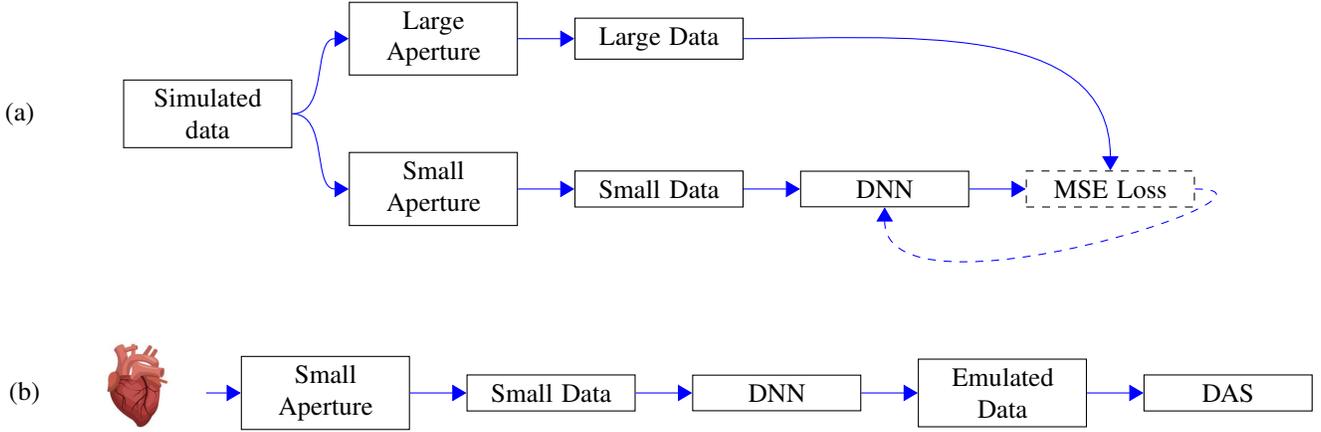}};
  
  \node[draw, text width=2cm,align=center] (DNN_node22) at (-4,0) {Small Aperture};
  \node[draw, text width=2cm,align=center] (DNN_node23) at (-1,0) {Small Data};
\node[draw, text width=2cm,align=center] (DNN_node24) at (2,0) {DNN};
  \node[draw, text width=2cm,align=center] (DNN_node25) at (5,0) {Emulated Data};
\node[draw, text width=2cm,align=center] (DNN_node26) at (8,0) {DAS};
  % CNN out
  \draw[-{Triangle[scale=1.5]},draw=blue] (DNN_node1) to[in=180,out=0] (DNN_node22);
  \draw[-{Triangle[scale=1.5]},draw=blue] (DNN_node22) to (DNN_node23);
\draw[-{Triangle[scale=1.5]},draw=blue] (DNN_node23) to (DNN_node24);
\draw[-{Triangle[scale=1.5]},draw=blue] (DNN_node24) to (DNN_node25);
\draw[-{Triangle[scale=1.5]},draw=blue] (DNN_node25) to (DNN_node26);

    \end{tikzpicture}

\caption{The setup for the two stages of our method: (a) The network training. (b) The network utilization.}
\label{fig:conceptual_diagram}
\end{figure*}

\section{SIGNAL MODEL}

\subsection{Synthetic Aperture Beamforming}

The full mathematical background for using the SA technique is given in \cite%
{norton1980reconstruction}, and a summary is provided here for the benefit
of the reader.

We consider a two-dimensional reflecting medium, denoted by $f(x,z)$ in the $%
x-z$ plane and the omnidirectional transducer located at $x_{0}$ on the
boundary (see Figure~\ref{fig:fxz}). Suppose a narrow pulse is transmitted
by the transducer, propagates isotropically in the $x-z$ plane. Following
the transmission of that pulse, the transducer is switched to a receiving mode
and the returning reflections are recorded as a function of time. Let $%
\mathcal{C}(x_{0},r)$ denote the semicircular path of radius $r$ and center
at $(x=x_{0},z=0)$. After a further delay of $r/c$ (where $c$ is the speed
of sound), the returned pulses originating from points lying along this
path, arrive simultaneously at the receiver and are integrated there. The
transducer is then shifted to a new location on the $x$ axis and the
acquisition process is repeated. This way, we generate the two-dimensional
(data) function $g(x_{0},r)$, where the connection between $f(x,z)$ and $%
g(x_{0},r)$ is given by 
\begin{align} \label{eq:41}
&g_{SA}\left( x_{0},r\right) =\oint_{\mathcal{C}\left( x_{0},r\right)
}f(x,z)dl \\ \nonumber
&=\int_{-\infty }^{\infty }\int_{-\infty }^{\infty }f\left( x,z\right) \delta\left( r-\left[ z^{2}+\left( x_{0}-x\right) ^{2}\right] ^{\frac{1}{2}%
}\right) dxdz
\end{align}%
which denotes the line integral of $f(x,z)$ along the semicircular path $%
\mathcal{C}(x_{0},r)$. Note that the transmitted signal is assumed to be the
Dirac Delta. The goal is to express the original function $f(x,z)$
in terms of $g(x_{0},r)$. Norton managed to derive an analytic expression
for this in a form of \emph{deconvolution}. We will further discuss some
aspects of his work in the sequel.

Clearly, a more efficient utilization of an array could be achieved by transmitting
from one element and receiving with all transducers. This is the STA
technique, presented next.

\subsection{Synthetic Transmit Aperture Beamforming}

A generalization of the SA technique is the Synthetic Transmit Aperture
(STA) \cite{jensen2006synthetic}. Similar to SA, a single element is used in
transmission, whereas in receive, all elements are used and a low-resolution
image can be formed. After the first element is used to create the first
low-resolution image, the successive element transmits to create another
low-resolution image. Once all the elements are used in transmission, the
low-resolution images are compounded to form the final high-resolution
image. As opposed to SA, STA imaging has the advantage of lower sidelobe
level and a higher SNR ratio. The data generation process can now be modeled
by 
\begin{align} \label{eq:42}
&g_{STA}\left( x_{t},x_{r},r\right)  =\oint_{\mathcal{C}\left(
x_{t},x_{r},r\right) }f(x,z)dl \\ \nonumber
& =\int_{-\infty }^{\infty }\int_{-\infty }^{\infty }f(x,z)  \delta \left( r-%
\frac{1}{2}\left( \left[ z^{2}+\left( x_{t}-x\right) ^{2}\right] ^{\frac{1}{2%
}}\right. \right.  \\ \nonumber
& +\left. \left. \left[ z^{2}+\left( x_{r}-x\right) ^{2}\right] ^{\frac{1}{2}%
}\right) \right) dxdz  \notag
\end{align}%
where $x_{t}$ and $x_{r}$ are the locations of transmitting and receiving
elements respectively, and $r(=t/c)$ represents the time passed from
transmission. In this case, we note that the recorded data is the result of
integration of reflected signal from points $\left( x,z\right) $\ along the
line $\mathcal{C}\left( x_{t},x_{r},r\right) $ which is half an ellipse with
focal points at $x_{t}$, $x_{r}$ and $r$ is the sum of distances to these
two focal points (see Figure~\ref{fig:fxz_STA}). The result is a three
dimensional data function, $g\left( x_{t},x_{r},r\right) $. This model is
clearly more complex than the SA case and to our knowledge no analytical
result for the reconstruction, similar to the one for the SA case, is
available. However, we wish to point out that since clearly, $g_{SA}\left(
x_{0},r\right) =g_{STA}\left( x_{0},x_{0},\frac{r}{2}\right) $, the exact
reconstruction of the image from data in the STA case should be possible as
well.

\subsection{Conventional Phased Array Beamforming}

Whereas in SA/STA techniques only a single element is active during the
transmission phase, in conventional Phased Array imaging all array elements
are used during both transmit and receive. By using all the elements simultaneously at the transmit phase, a beam is created and electronically swept
over the region of interest (ROI) in the image.

\begin{figure*}[t!]
\centering
\begin{tikzpicture}
  % CNN in
  \node[inner sep=0pt] (DNN_node1) at (-13,0) {\includegraphics[width=0.35\textwidth]{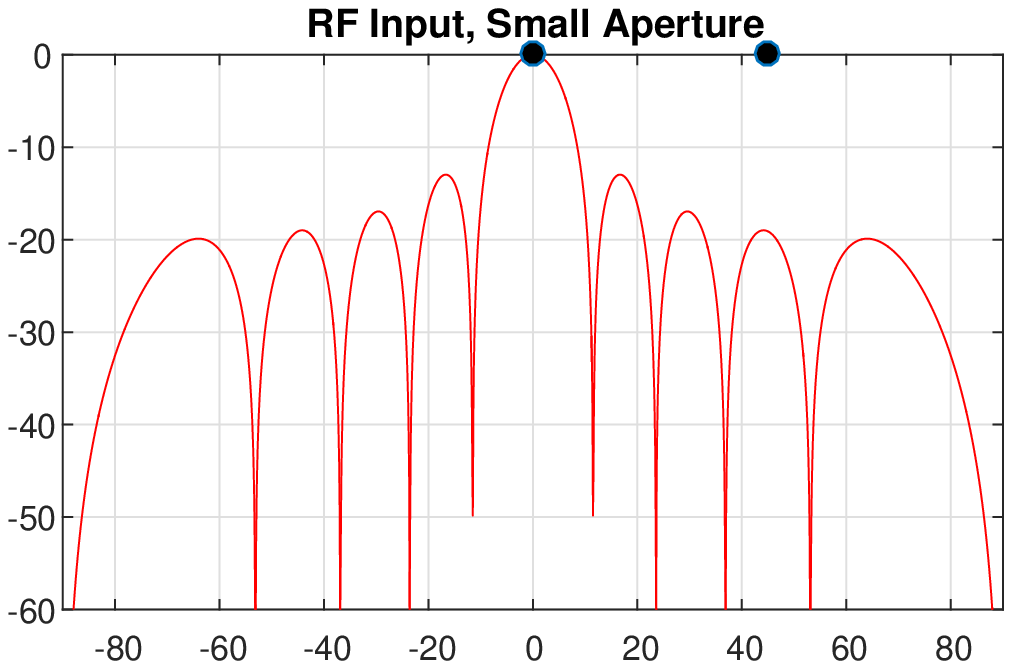}};
  \node[draw, text width=1.5cm,align=center] (DNN_node2) at (-8,0) {DNN};
  \node[inner sep=0pt] (DNN_node3) at (-3,0) {\includegraphics[width=0.35\textwidth]{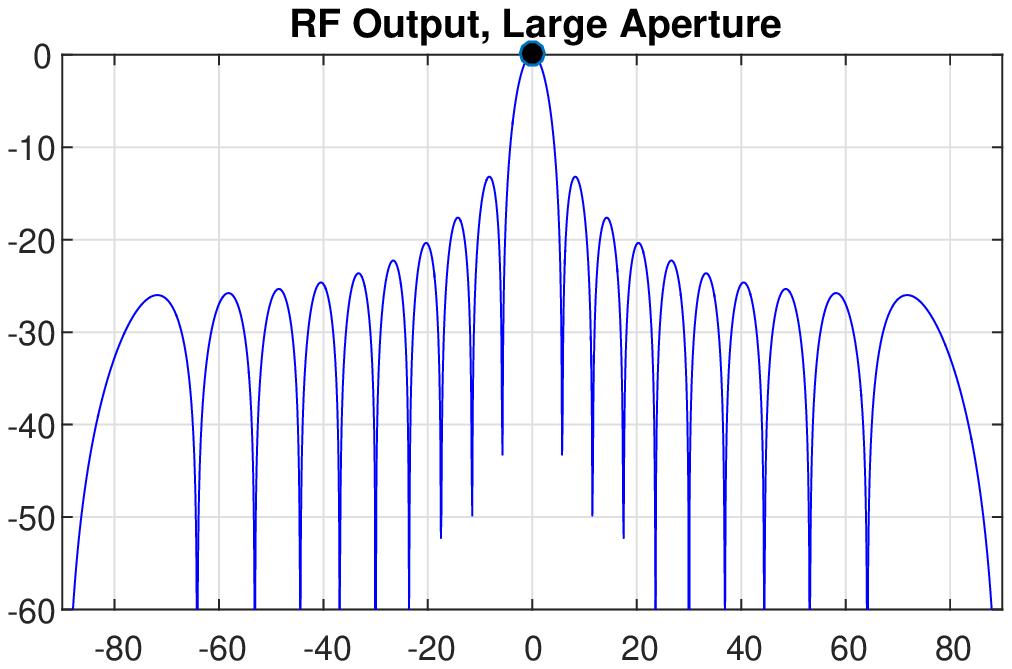}};
  % CNN in
  % CNN out
  \draw[-{Triangle[scale=1.5]},draw=blue] (DNN_node1) to (DNN_node2);
  \draw[-{Triangle[scale=1.5]},draw=blue] (DNN_node2) to (DNN_node3);
    \end{tikzpicture}
\caption{An illustration of the proposed training method. Left: receive
beam-pattern of the small aperture (red). Right: receive beam-pattern of the
large aperture (blue). Black dots represent point targets.}
\label{fig:training_process}
\end{figure*}

\section{DEEP LEARNING BASED BEAMFORMING}

As mentioned earlier, in his work, \cite{norton1980reconstruction} and \cite{norton1977theory}, Norton derived an analytical inverse mapping of Eqn. %
(\ref{eq:41}). He further derived an analytic expression for the result of
applying the DAS operation on the SA generated data. While his analysis was
based on the assumption of an infinite array, in \cite{norton1977theory} we find
an analysis and discussion of how having a finite array affects the image
reconstruction results both for the deconvolution and the DAS methods. The
observations made by Norton provided the motivation for our work.
Specifically:

\begin{enumerate}
\item The reconstruction using DAS is only an attempt to approximate the
exact reconstruction, the deconvolution. Namely, since both use the same
data, the DAS does \emph{not fully utilize the data collected by the array.}

\item Constrained to a finite size array deteriorates the quality of the
reconstructed image whether using the deconvolution or the DAS. The smaller
the array, the worse the reconstructed image quality. However, \emph{the
impact on the DAS result is more significant}.
\end{enumerate}

With these two observations and inspired by \cite{dong2015image} who applied
deep learning to get super resolution from a single image, we have developed
the method we present here. Given data from a finite size aperture scan of
an image, we train a neural network to emulate data resulting from scanning
the same image with a \emph{larger aperture}. Then, applying DAS on the
larger (emulated) data we generate a considerably improved reconstructed
image. This result is directly related to a fundamental result in array
signal processing: Increasing array size while keeping the same
inter-element spacing, achieves a narrower main lobe (better lateral
resolution), and lower sidelobes (better contrast) relative to a small array
size.

In Figure~\ref{fig:conceptual_diagram} we present the setup for the two
stages of our method: The network training and the network utilization.

 \begin{figure*}[h]
    \centering
    \includegraphics[width=\textwidth]{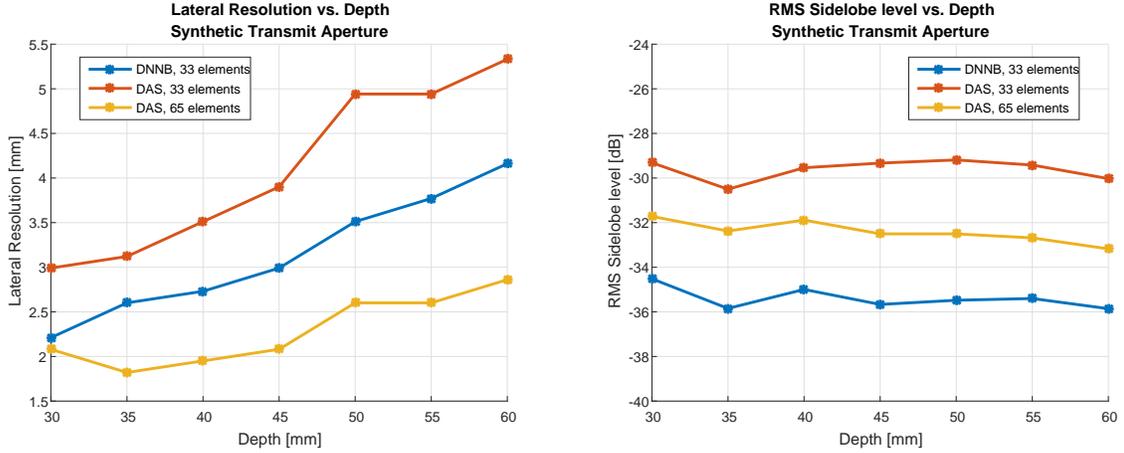}
    \caption{Left: STA lateral resolution vs. depth. Right: STA RMS sidelobe level vs. depth.}
    \label{fig:STA_rms_sll_lateral_res}
\end{figure*}

In our experiments with neural networks, we managed to achieve a further
improvement of contrast resolution using a neural network trained to reduce
sidelobes. To this end, in addition to its emulation role, we also train the
network to place deeper nulls in the sidelobes locations. To achieve this,
the input training example contains a multichannel waveform as received from
two targets: one in the main lobe and another in a random sidelobe. The
output training example contains a multichannel waveform as received from just
a single target in the main lobe. An illustration of this proposed training
process is given in Figure~\ref{fig:training_process}. 

We draw some similarity to our problem from \cite{luchies2018deep}, who
trained a network to suppress off-axis scattering outside the first nulls of
the beam. However, there are significant differences between our
approaches. First, \cite{luchies2018deep} operates in the frequency domain
via the short-time Fourier transform, where different networks were trained
for different frequencies. We propose to operate in the space/time domain,
training a single network. Second, we use two separate arrays of different
sizes for the data generation stage, whereas \cite{luchies2018deep} uses a
single array whose size is fixed. Finally, as pointed out earlier, we are
also aiming at providing an improvement in lateral resolution of the
reconstructed image.

\section{Experiments}
The experiments we describe in the sequel are divided into two parts. In the
first we used simulated data while in the second we used real cardiac data.
In both parts, we compare the image quality for a given (small) array,
using DAS and the DNNB. Our method clearly results in significantly superior
image quality. An interesting question is how close is the data emulated by
the DNN to real data generated by the large array. Using simulated data, we
could also compare the performance of our trained DNN with DAS applied to
data generated by the large array. This can be observed in Table \ref{table:All_table} and Figure~\ref{fig:STA_rms_sll_lateral_res}.

We also wish to point out that with the real data experiments, we were
constrained by the available data. So, since all this data was collected by
a given array, both transmit and receive, we experimented only with the
possibility of a small array in receive where small array simply means using
the data from a subset of receiving elements.

\subsection{Training The Network}
Training the network in a supervised manner requires a training data set
with a large number of low/high-resolution multichannel waveform pairs. Due
to practical constraints, we used in our experiments for the training stage
of the DNN only data simulated by FIELD II \cite{jensen1996field}. The data
set comprised pairs of 30,000 point targets randomly positioned in various
DOA angles, simulated by a uniform linear array of 33 elements
(low-resolution signals), and the corresponding point targets simulated by a
uniform linear array of 65 elements (high-resolution signal). We note that
our choice of the ratio $65/33\approx 2$ seems quite arbitrary but in our
experiments seemed to produce good results. 

The synthetic data generation enabled us to train simultaneously the network
for both the emulation part and the sidelobe suppression part as explained
earlier. Given a set of high-resolution multichannel waveform matrices, and
their corresponding low-resolution multichannel waveform matrices, we train
the network to minimize the mean-squared error (MSE) loss function (see
Figure \ref{fig:conceptual_diagram}).

Clearly, for each of the scanning techniques we experimented with, SA, STA and PA, a
separate DNN had to be trained. In the conventional PA method, all scan lines
typically have a fixed origin on the transducer surface, but are steered in
a fan pattern to create the image. We delay each channel in the training
data set to focus at a point in the image, which is the only pre-processing
we perform on the data before being fed into the DNN.

\begin{table}[t!]
    \centering
    \caption{Field II simulation parameters}
    \begin{tabular}{|c|c|c|c|} 
    \hline
    \textbf{Parameter} & \textbf{Value} \\
    \hline
    c (Speed of Sound) & 1540 [m/sec] \\ 
    $f_{0}$ (Central Frequency) & 3.5 [MHz] \\ 
    $f_{s}$ (Sampling Frequency) & 16 [MHz] \\ 
    Element Width & 0.220 [mm] \\ 
    Element Height & 5 [mm] \\ 
    Kerf & 0.044 [mm] \\ 
    Number of Elements & 33/65 \\ 
    Apodization & None \\ 
    Number of Scan Lines & 65 \\ 
    Sector Size & 48$\degree$ \\ 
    Tx Focus & 50 [mm] \\ 
    \hline
    \end{tabular}
    \label{table:Field_table}
\end{table}

\textbf{Architecture:} We used a fully connected neural network, followed by convolutional layers. The number of hidden layers was 6. The Leaky Rectified Linear Unit (LeakyReLU) was chosen as the activation function of hidden layers, where the last layer was linear. The full network effectively scales the input multichannel waveform dimensions up by a factor 2. We used Keras API \cite{chollet2015keras} on top of a TensorFlow \cite{tensorflow2015} backend written in Python to create and train the network. \par

\textbf{Optimization:} The network was trained for 50 epochs using the Adam optimizer \cite{kingma2014adam} with a learning rate of $10^{-3}$, and learning rate decay of $10^{-8}$. As mentioned earlier, we used the MSE as the loss function. We reduced the learning rate by a factor of 2 when the validation loss has stopped improving. Weights of all neurons in the network were initialized using the Uniform distribution. The mini-batch size was chosen to be 64 samples to attain a good trade-off between the efficiency of not having all training data in memory, and model convergence speed. \par

\subsection{Simulation Setup}
To evaluate our proposed beamformer, we first tested it on synthetic cyst phantom, and then on a set of consecutive frames of cardiac ultrasound data. The cyst phantom scan was done using an aperture comprising 33/65 transducer elements, operating with a central frequency of $f_0 = 3.5 [MHz]$. The width of each element, measured along the \text{\^{x}} axis, was $\frac{1}{2}\lambda = \frac{c}{2f_0} = 0.220[mm]$, and the height, measured along the \text{\^{y}} axis, was $5[mm]$. The elements were arranged along the \text{\^{x}} axis, with a $0.044[mm]$ inter-element spacing (kerf). The transmitted pulse was simulated by exciting each element with $1.75$ periods of a sinusoid of frequency $f_0$, where the delays were adjusted such that the transmission focal point was at depth $r = 50[mm]$. No apodization was applied on transmission (rectangular window).\par

The RF signals of each channel were sampled at $16[MHz]$. During the simulations, a $48\degree$ sector was imaged using 65 scan lines. The envelope was extracted using the Hilbert transform, and the appropriate beamforming was applied according to the selected technique. The simulation parameter settings are summarized in Table \ref{table:Field_table}.

\subsection{Quality Measures}
In this section, we describe the quantitative measurements used to assess the proposed method, DNNB, performance in terms of lateral resolution and sidelobes level. The lateral resolution is evaluated by computing the full width at half maximum of the point spread function in the lateral axis. A quantitative measure of contrast can be calculated by the contrast-to-noise ratio (CNR) and contrast ratio (CR):
\begin{align*}
\mathbf{CNR}=\frac{\left\vert\mu_{\mathrm{b}}-\mu_{\mathrm{c}}\right\vert}{\sqrt{\sigma_{\mathrm{b}}^{2}+\sigma_{\mathrm{c}}^{2}}}
\end{align*}
\begin{align*}
  \mathbf{CR}=20\log_{10}{\left(\frac{\mu_{\mathrm{c}}}{\mu_{\mathrm{b}}}\right)}
\end{align*}
where $\mu_{\mathrm{c}}$ and $\mu_{\mathrm{b}}$ are the mean image intensities, respectively in a region inside the phantom cyst or in the surrounding background, and $\sigma_{\mathrm{c}}^{2}$, $\sigma_{\mathrm{b}}^{2}$ are the corresponding variances \cite{lediju2011short}. A quantitative measure of sidelobes level is the root mean square (RMS) sidelobes level, which is evaluated by computing the intensity of the point spread function outside the mainlobe.

%%%

\section{Results}
\subsection{Simulated Data}
Next, we present the results of our experiments obtained by applying DNNB to
ultrasound signals simulated using the Field II program for an image of cyst
phantom. 

\subsubsection{SA Results}
\begin{figure}[t]
    \centering
    \includegraphics[width=0.5\textwidth, height=0.2\textwidth]{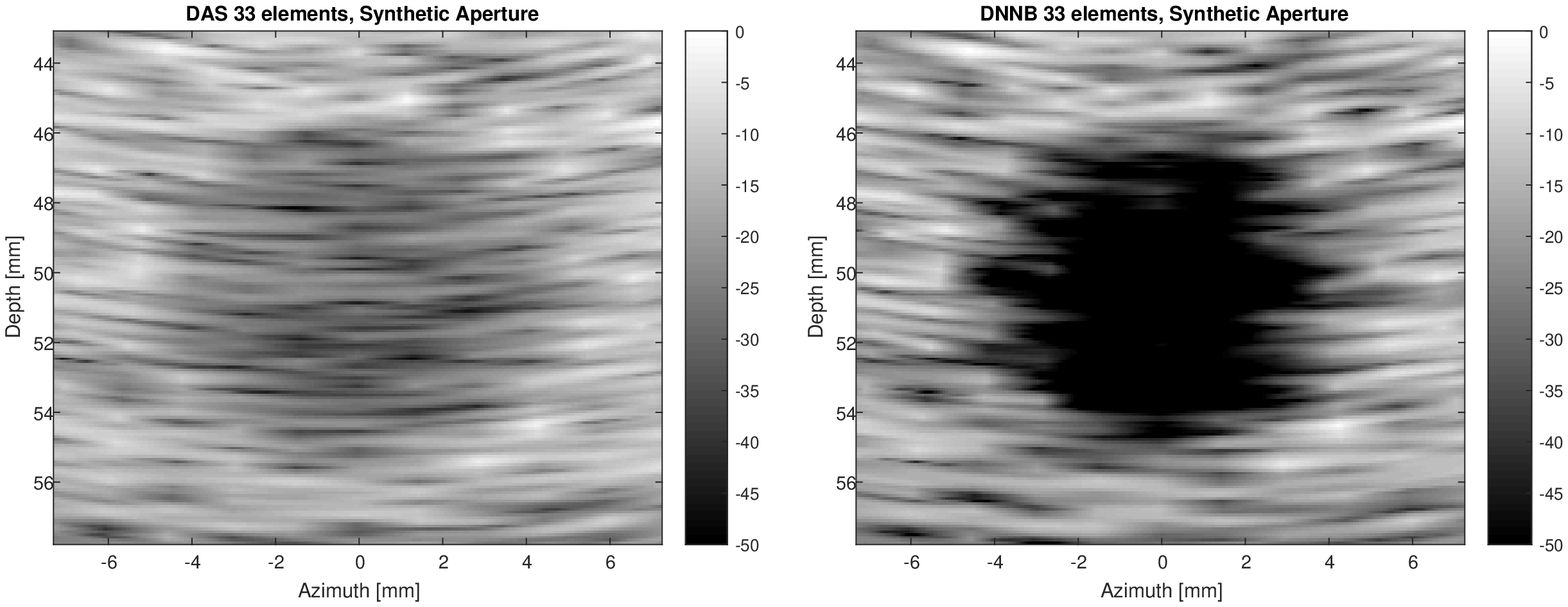}
    \caption{SA results of standard DAS and the proposed DNN for phantom data, using an array of 33 elements.}
    \label{fig:SA_cyst_das33_dnn33_vertical}

    \centering
    \includegraphics[width=0.5\textwidth, height=0.2\textwidth]{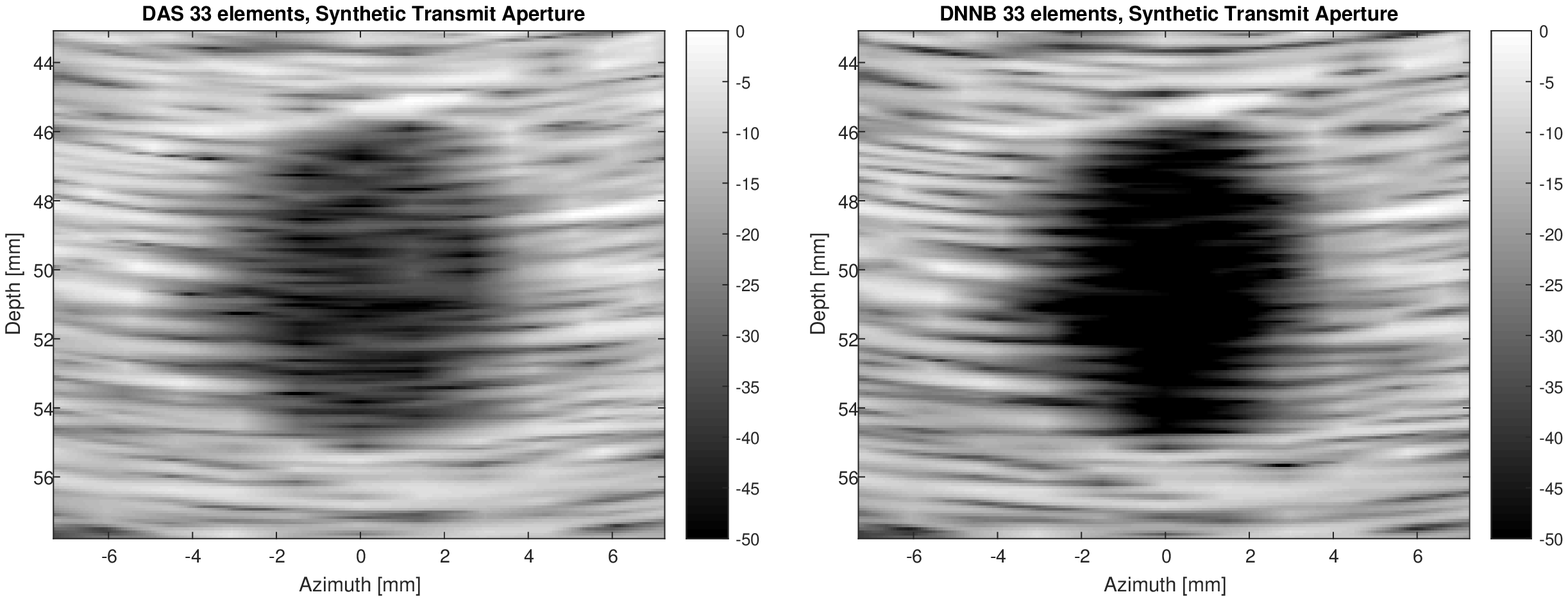}
    \caption{STA results of standard DAS and the proposed DNN for phantom data, using an array of 33 elements.}
    \label{fig:STA_cyst_das33_dnn33_vertical}

    \centering
    \includegraphics[width=0.5\textwidth, height=0.2\textwidth]{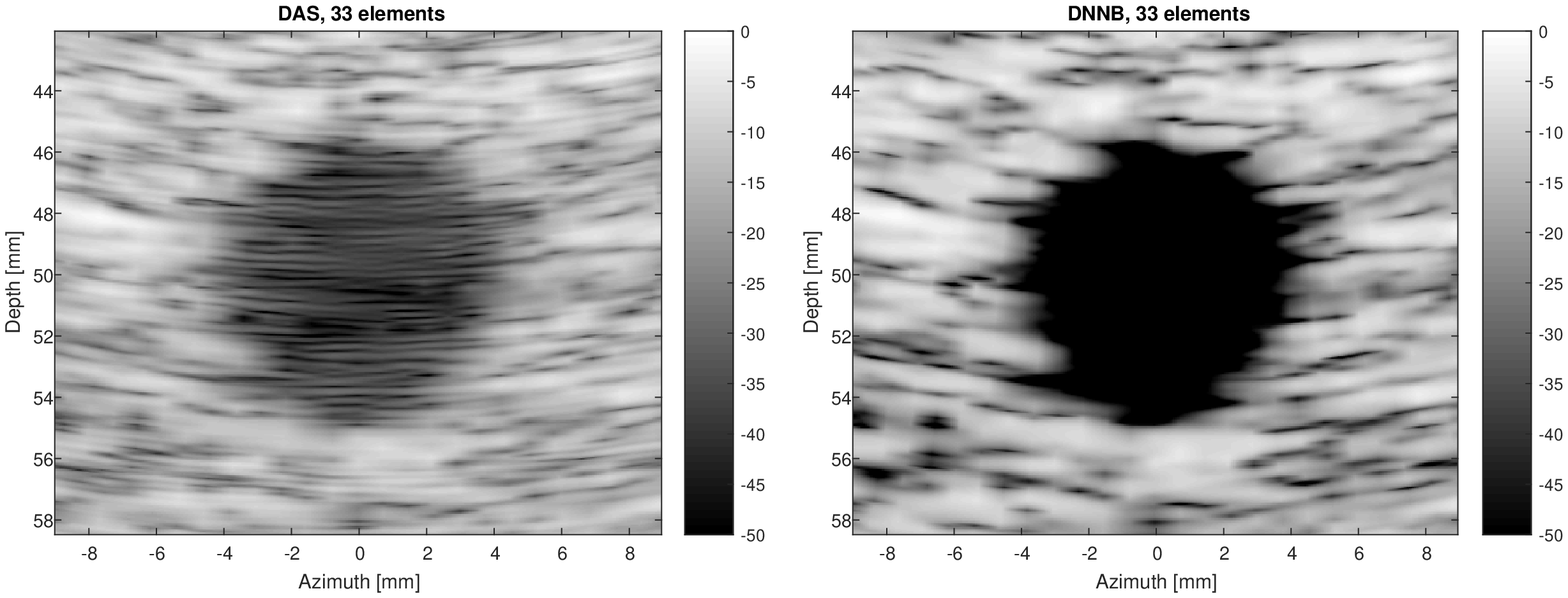}
    \caption{PA results of standard DAS and the proposed DNN for phantom data, using an array of 33 elements.}
    \label{fig:cyst_das_33_65_dnn}
\end{figure}

As can be seen in Figure~\ref{fig:SA_cyst_das33_dnn33_vertical}, the main
effect of the DNNB is in the reduction of the lateral sidelobes, which is
expressed in contrast enhancement: the CR is -5.7779 [dB] compared to
-3.0151 [dB] with DAS 33 elements. The CNR is also improved, 0.0296,
compared to 0.0165 with DAS 33 elements. In addition, it can be seen in
Figure~\ref{fig:SA_cyst_das33_dnn33_vertical} that DNNB provides better
sidelobe interference reduction within the cyst, while preserving the
speckle texture in the region outside the cyst. This provides empirical
justification for the proposed training method, where we trained the network
to place deeper nulls at the sidelobes.

\subsubsection{STA Results}
We next compare the results of a cyst phantom, obtained by Synthetic
Transmit Aperture imaging (see Figure~\ref{fig:STA_cyst_das33_dnn33_vertical}%
). The contrast of the anechoic cyst is improved when applying DNNB: the CR
is -7.4324 [dB] compared to -4.9211 [dB] with DAS 33 elements. The CNR is
also improved: 0.0362, compared to 0.0259 with DAS 33 elements. We also
study the robustness of DNNB to depth changes. To that end, point targets
were placed at various depths. A graph of STA performance as a function of
depth is presented in Figure~\ref{fig:STA_rms_sll_lateral_res}.

%%%%%%%%%%%%%%%%% PA %%%%%%%%%%%%%%%%%%%%%%%%%
\subsubsection{Conventional Phased Array imaging Results}
We next proceed to the results of a cyst phantom obtained by Conventional
Phased Array (PA) imaging. The image produced by DNNB is characterized by a
very sharp transition from speckle to cyst region, noted in Figure~\ref%
{fig:cyst_das_33_65_dnn}. As expected, the contrast of the anechoic cyst is
improved when applying DNNB: the CR is -10.8735 [dB] while it is -5.2532
[dB] with DAS 33 elements. The CNR is also improved: 0.0591, while it is
0.0343 with DAS 33 elements.

The resulting values of lateral resolution and RMS sidelobe level of a single point target are summarized in Table \ref{table:All_table}. It can be noted that DNNB 33 elements provide a prominent improvement in lateral resolution and sidelobes level compared to standard DAS 33 elements processing.\par

\subsection{Cardiac Data Experiments}
\begin{figure}[h]
    \centering
    \includegraphics[width=0.55\textwidth]{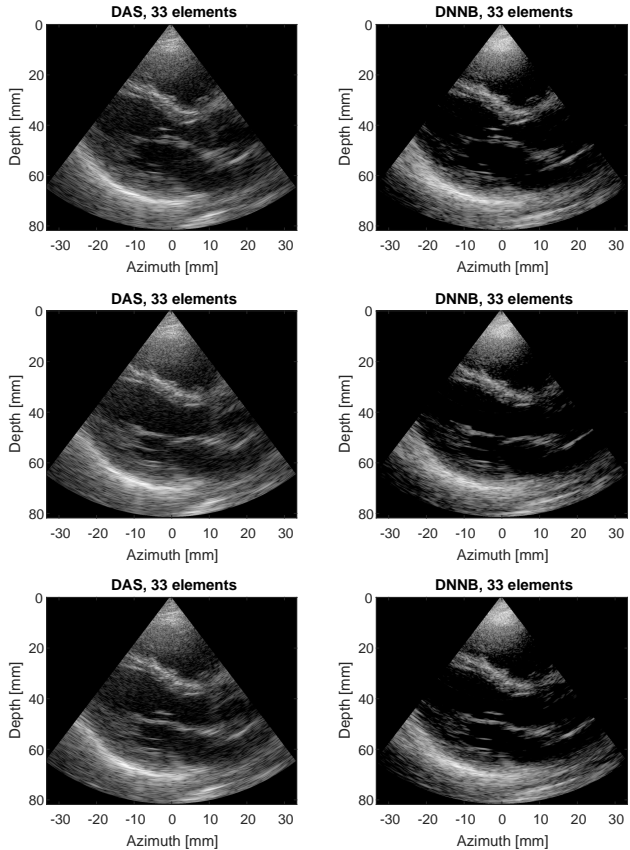}
    \caption{Real Cardiac data results. The first column corresponds to DAS, and the second corresponds to the proposed method. Each row is corresponding to a different frame.}
    \label{fig:card_das_33_dnn_all_frames_no_titles}
\end{figure}

We applied the DNNB on real cardiac ultrasound data. This allows for a (partial) qualitative assessment of the proposed technique. The acquisition was performed using a wideband phased array ultrasound scanner consist of 64 elements. Operating in second harmonic imaging mode, pulses were transmitted at $1.7 [MHz]$, and the corresponding second harmonic signal, centered at $3.42 [MHz]$, was then acquired. Data from all acquisition channels were sampled at $20 [MHz]$ and collected along 120 beams, forming a $75\degree$ sector. The beamforming results obtained for the cardiac data are depicted in Figure~\ref{fig:card_das_33_dnn_all_frames_no_titles}. 

\subsection{Cardiac Data Experiments - Receive Aperture Reduction}
\begin{figure*}[h]
    \centering
    \includegraphics[width=\textwidth]{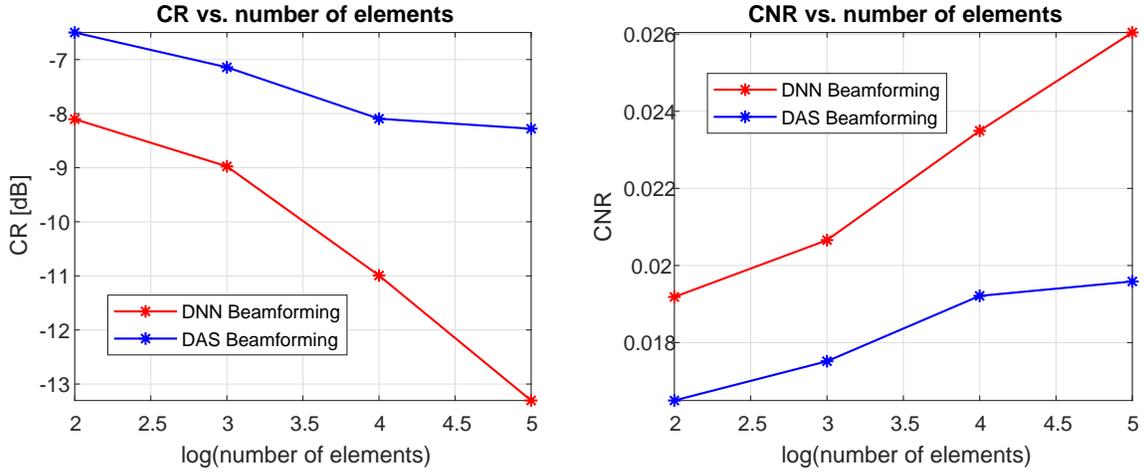}
    \caption{Left: Contrat ratio vs. number of elements. Right: Contrst-to-noise ratio vs. number of elements.}
    \label{fig:CR_CNR_vs_num_of_elem}
\end{figure*}

\begin{figure*}[h]
\begin{subfigure}{0.5\textwidth}
\includegraphics[width=1\textwidth]{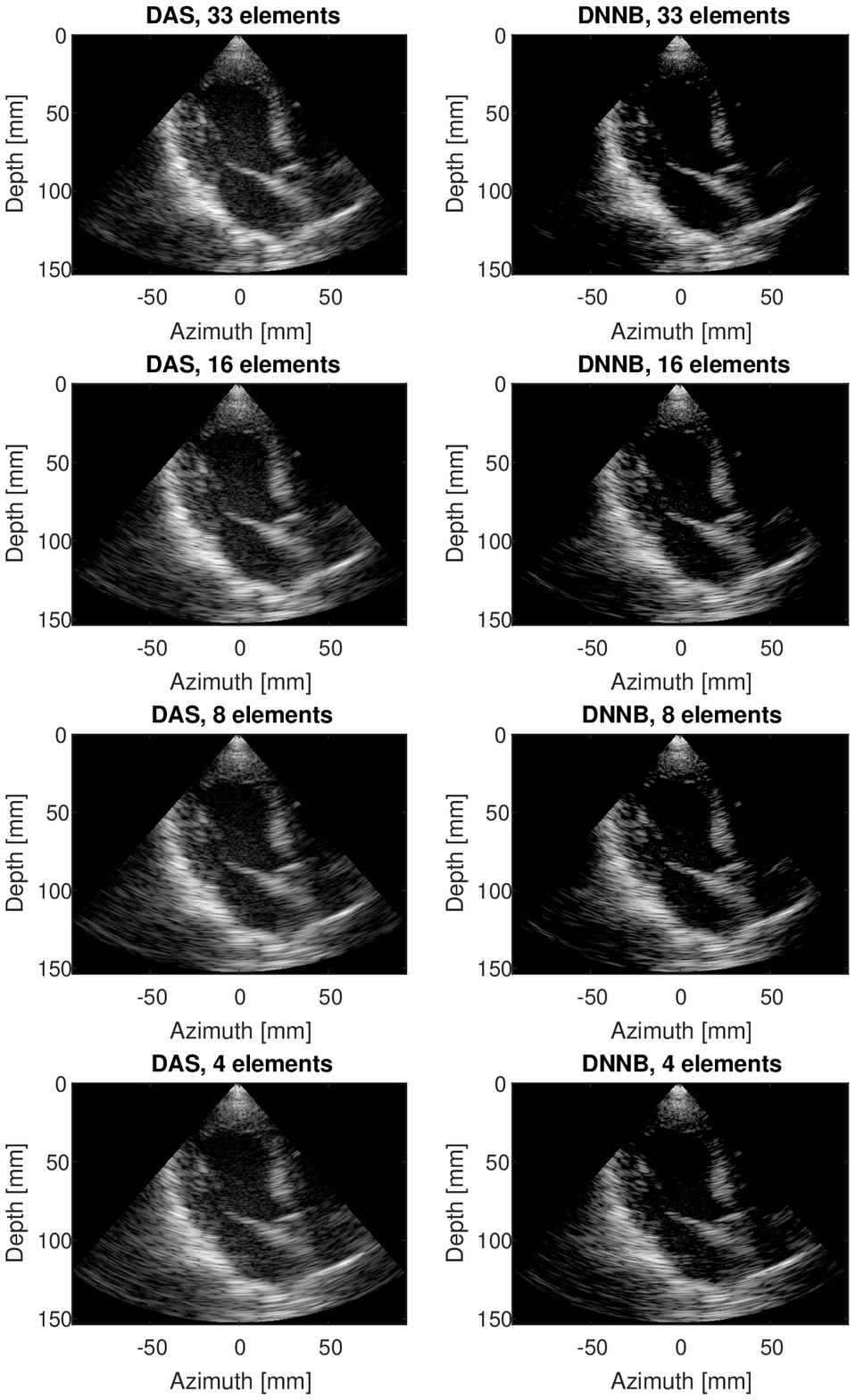}
\caption{}
\label{fig:card_sparse_array}
\end{subfigure}
\begin{subfigure}{0.5\textwidth}
\includegraphics[width=1\textwidth]{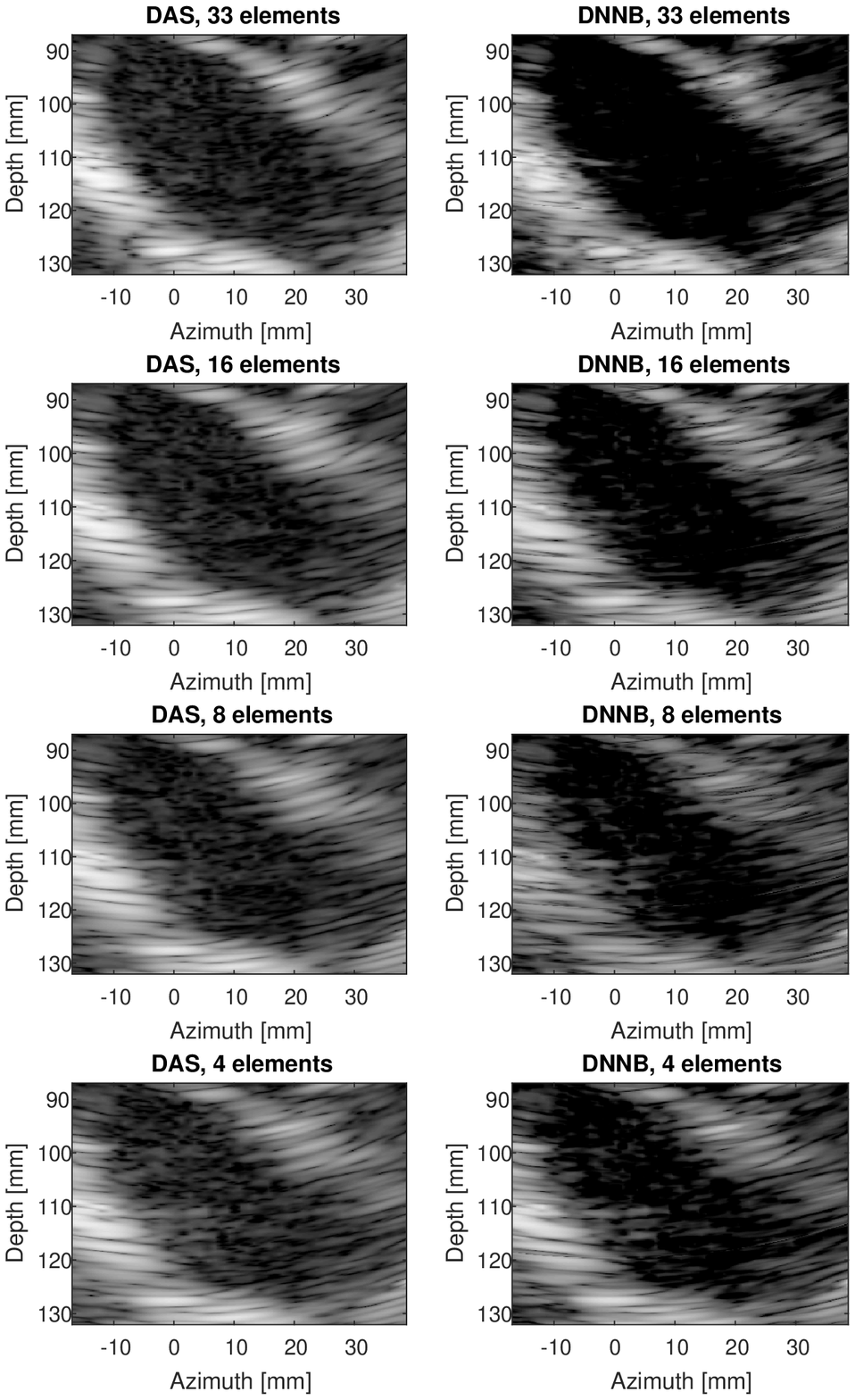}
    \caption{}
    \label{fig:card_sparse_array_zoom_in}
\end{subfigure}
\caption{(a) Real Cardiac data results. The first column corresponds to DAS, and the second corresponds to the proposed method. Each row is corresponding to a different reduction factor. (b) Zoom in of (a).}
\label{fig:image2}
\end{figure*}

We used \textit{in-vivo} acquisitions from \cite{rindal2017hypothesis} to study the robustness of the proposed beamformer to 
different aperture sizes, where the receive aperture was reduced by a factor of 2, 4, and 8. The results are illustrated in Figure~\ref{fig:card_sparse_array}, where each row corresponds to a different reduction factor. When looking at the magnified images (see Figure~\ref{fig:card_sparse_array_zoom_in}) it is clearly observed that the sidelobes reflections, detected in the middle of the left ventricle, is indeed locally weaker in magnitude compared with the speckle reflections in the surrounding. A logarithmic scale graph of CR and CNR as a function of the number of elements is presented in Figure~\ref{fig:CR_CNR_vs_num_of_elem}.

\begin{table}[t]
    \centering
    \caption{SA, STA and Phased Array results for lateral resolution and RMS sidelobe level.}
    \begin{tabularx}{0.4\textwidth} { 
      | l 
      | >{\centering\arraybackslash}X 
      | >{\centering\arraybackslash}X | }
     \hline
     \textbf{Method} & \textbf{Lateral Resolution [mm]} & \textbf{RMS sidelobe level [dB]}\\
     \hline
     SA DAS 65 elements & 1.951 & -30.95   \\
     \hline
     SA DAS 33 elements & 3.961 & -27.71  \\
     \hline
     SA DNNB 33 elements & 2.341 & -32.34  \\
    \hline
     STA DAS 65 elements & 2.601 & -32.51   \\
     \hline
     STA DAS 33 elements & 4.871 & -29.19  \\
     \hline
     STA DNNB 33 elements & 3.511 & -35.48  \\
    \hline
     PA DAS 65 elements & 2.812 & -36.23   \\
     \hline
     PA DAS 33 elements & 4.941 & -31.76 \\
     \hline
     PA DNNB 33 elements & 3.774 & -39.12  \\
    \hline
    \end{tabularx}
        \label{table:All_table}
\end{table}

\section{Conclusion}
In this paper, we proposed a deep learning-based method, DNNB, for the beamforming
process in medical ultrasound. The network is trained to achieve two
purposes: One, to emulate data from a larger array, and two, to suppress
sidelobes. By this, we managed to achieve both an improved resolution and
improved contrast from a given array over the use of DAS on the available
data. We demonstrate this both by simulation results and experiments on real
cardiac data. The comparison shows the superior performance of DNNB in terms
of both lateral resolution and contrast resolution. 
% trade off between contrast and lateral resolution
Our method shows the ability of sidelobe suppression using a neural network with fewer elements. The performance comparison between DAS 65 elements and DNNB 33 elements points to a trade-off between lateral resolution and contrast: the main improvement introduced by the proposed method is the significantly better contrast resolution. We note, however, that there is also better lateral resolution compared to DAS with the same number of elements. The reconstructed images obtained with the DNN beamformer achieve better CR and CNR due to significant suppression of the RMS sidelobe level compared to DAS. Accordingly, the reconstructed phantom cyst shows an improved edge definition, while preserving the speckle texture in the surrounding. \par

In conclusion, DNNB out performs the DAS method when applied to the same data. Consequently, as pointed out earlier, fewer elements are required to reconstruct an image with high contrast and lateral resolution, paving the way to more efficient ultrasound devices. \par

%\section*{Acknowledgment}
%The authors would like to thank...

% Can use something like this to put references on a page
% by themselves when using endfloat and the captionsoff option.
\ifCLASSOPTIONcaptionsoff
  
\fi

% trigger a  just before the given reference
% number - used to balance the columns on the last page
% adjust value as needed - may need to be readjusted if
% the document is modified later
%\IEEEtriggeratref{8}
% The "triggered" command can be changed if desired:
%\IEEEtriggercmd{\enlargethispage{-5in}}

\bibliographystyle{IEEEtran} 
\bibliography{general}

% biography section
% 
% If you have an EPS/PDF photo (graphicx package needed) extra braces are
% needed around the contents of the optional argument to biography to prevent
% the LaTeX parser from getting confused when it sees the complicated
% \includegraphics command within an optional argument. (You could create
% your own custom macro containing the \includegraphics command to make things
% simpler here.)
%\begin{IEEEbiography}[{\includegraphics[width=1in,height=1.25in,clip,keepaspectratio]{mshell}}]{Michael Shell}

\begin{IEEEbiographynophoto}{Nissim Peretz}
received the B.Sc. degree in electrical engineering from the Department of Electrical Engineering, Technion, Haifa, Israel, in 2015, where he is currently pursuing the M.Sc. degree in Electrical Engineering at the Technion.
\end{IEEEbiographynophoto}

\begin{IEEEbiographynophoto}{Arie Feuer}
(M’76–SM’93–F’04–LF’14) has been with the Department of Electrical Engineering at the Technion - Israel Institute of Technology, since 1983 where he is currently a Professor Emeritus. He received his B.Sc. and M.Sc. from the Technion in 1967 and 1973 respectively and his Ph.D. from Yale University in 1978. From 1967 to 1970 he was in industry working on automation design and between 1978 and 1983 with Bell Labs in Holmdel, N.J. Between the years 2013 and 2015 he worked for a startup company developing algorithms for medical ultrasound imaging. Between the years 1994 and 2002 he served as the president of the Israel Association of Automatic Control and was a member of the IFAC Council during the years 2002-2005. Arie Feuer is a Life Fellow of the IEEE. In the last 25 years he has been regularly visiting the EECS department at the University of Newcastle. In 2009 he received an honorary doctorate from the University of Newcastle. His current research interests include medical imaging, in particular ultrasound and CT, resolution enhancement of digital images and videos, 3D video and multi-view data, sampling and combined representations of signals and images, and adaptive systems in signal processing and control.
\end{IEEEbiographynophoto}

% You can push biographies down or up by placing
% a \vfill before or after them.
\vfill

% Can be used to pull up biographies so that the bottom of the last one
% is flush with the other column.
%\enlargethispage{-5in}

\end{document}